\documentclass[MSNbibl,nameyear,dvips]{arxstspdf}
\usepackage{multirow}
\usepackage{graphicx}
\usepackage{flushend}
\usepackage{stfloats}

\volume{27}
\issue{2}
\pubyear{2012}
\firstpage{232}
\lastpage{246}
\doi{10.1214/11-STS369}

\makeatletter

\setattribute{abstract}{width}{345pt}
\setattribute{keyword}{width}{345pt}

\newcommand{\bb}{\bolds{\beta}}
\newcommand{\by}{\mathbf{y}}
\newcommand{\bX}{\mathbf{X}}
\newcommand{\rr}{\mathcal{R}}
\newcommand{\xtx}[1]{\mathbf{X}_{#1}^T \mathbf{X}_{#1}}
\newcommand{\bsm} { \Sigma_m }
\makeatother

\begin{document}
\begin{frontmatter}

\title{Joint Specification of Model Space and Parameter Space Prior Distributions}

\runtitle{Prior Distributions}

\begin{aug}
\author[a]{\fnms{Petros} \snm{Dellaportas}\corref{}\ead[label=e1]{petros@aueb.gr}},
\author[b]{\fnms{Jonathan J.} \snm{Forster}\ead[label=e2]{J.J.Forster@soton.ac.uk}}
\and
\author[c]{\fnms{Ioannis} \snm{Ntzoufras}\ead[label=e3]{ntzoufras@aueb.gr}}

\runauthor{P. Dellaportas, J. J. Forster and I. Ntzoufras}

\affiliation{Athens
University of Economics and Business, University of
Southampton and University of Economics and Business}

\address[a]{Petros Dellaportas is Professor, Department of Statistics, Athens
University of Economics and Business, 10434 Athens, Greece \printead{e1}.}

\address[b]{Jonathan
J.~Forster is Professor, Southampton Statistical Sciences Research Institute,
University of
Southampton, Highfield, Southampton, SO17 1BJ, UK
\printead{e2}.}

\address[c]{Ioannis Ntzoufras is Associate Professor, Department of Statistics, Athens
University of Economics and Business, 10434 Athens, Greece \printead{e3}.}

\end{aug}

%
\begin{abstract}
We consider the specification of prior distributions for Bayes\-ian
model comparison, focusing on regression-type models. We
propose a particular joint specification of the prior
distribution across models so that sensitivity of
posterior model probabilities to the dispersion of prior
distributions for the parameters of individual models (Lindley's
paradox) is diminished. We illustrate the behavior of inferential and
predictive posterior quantities in linear and log-linear regressions
under our proposed prior densities with a series of simulated and real
data examples.
\end{abstract}

%

\vspace*{-6pt}
\end{frontmatter}

\section{Introduction and Motivation}\label{sec1}

A Bayesian approach to inference under model uncertainty proceeds as
follows. Suppose that the data~$\mathbf{y}$ are considered to have been
generated by a~model~$m$, one of a~set $M$ of competing models. Each
model specifies the distribution of $\mathbf{Y}$, $f(\mathbf{y} | m,\allowbreak
\bolds{\beta}_m)$ apart from an unknown parameter vector
$\bolds{\beta}_m\in B_m$, where~$B_m$ is the set of all possible values
for the coefficients of model $m$. We assume that $B_m=\rr^{d_m}$ where
$d_m$ is the dimensionality of $\bb_m$.

If $f(m)$ is the prior probability of model $m$, then the posterior
probability is given by
%
\begin{equation}\label{bmc}
f(m|\mathbf{y})={{{f(m)
f(\mathbf{y} |m)}}\over{\sum\limits_{m\in M} f(m)f(\mathbf{y}|m)}},
 \quad m\in M,\vadjust{\goodbreak}
\end{equation}
where $f(\mathbf{y}|m)$ is the marginal likelihood calculated using
$f(\mathbf{y}|m)=\int f(\mathbf{y}|m, \bolds{\beta}_m)
f(\bolds{\beta}_m|m)\,d\bolds{\beta}_m$ and\break $f(\bolds{\beta}_m| m)$ is
the conditional prior distribution of $\bolds{\beta}_m$, the model
parameters for model $m$. Therefore
\[
f(m|\mathbf{y})\propto{f(m) f(\mathbf{y} |m)},\quad  m\in M.
\]

For any two models $m_1$ and $m_2$, the ratio of the posterior model
probabilities (posterior odds in favor of $m_1$) is given
by
%
\begin{equation}\label{bf}\frac{f(m_1|\mathbf{y})}{f(m_2|\mathbf
{y})}=\frac{f(m_1)}{f(m_2)}\frac{f(\mathbf{y}|m_1)}{f(\mathbf{y}|m_2)},
\end{equation}
the
ratio of prior probabilities multiplied by the ratio of marginal
likelihoods, also known as the Bayes factor.

The posterior distribution for the parameters of a~particular model is
given by the familiar expression
\[
f(\bb_m|m, \mathbf{y})\propto f(\bb_m|m) f(\mathbf{y} |\bb_m, m),\quad
 m\in M.
\]
For a single model, a highly diffuse prior on the model parameters is
often used (perhaps to represent ignorance). Then the posterior density
takes the shape of the likelihood and is insensitive to the exact value
of the prior density function, provided that the prior is relatively
flat over the range of parameter values with nonnegligible likelihood.
When multiple models are being considered, however, the use of such a
prior may create an apparent difficulty. The most obvious manifestation
of this occurs when we are considering two models $m_1$ and~$m_2$ where
$m_1$ is completely specified (no unknown parameters) and $m_2$ has\vadjust{\goodbreak}
parameter $\bolds{\beta}_{m_2}$ and associated prior density
$f(\bolds{\beta}_{m_2}|m_2)$. Then, \textit{for any observed data
$\mathbf{y}$}, the Bayes factor in favor of $m_1$ can be made
arbitrarily large by choosing a sufficiently diffuse prior distribution
for $\bolds{\beta}_{m_2}$ (corresponding to a prior density
$f(\bolds{\beta}_{m_2}|m_2)$ which is sufficiently small over the range
of values of $\bolds{\beta}_{m_2}$ with nonnegligible likelihood).
Hence, under model uncertainty, two different diffuse prior
distributions for model parameters might lead to essentially the same
posterior distributions for those parameters, but very different Bayes
factors.

This result was discussed by Lindley (\citeyear{Lin57}) and is often
referred to as ``Lindley's paradox'' although it is also variously
attributed to Bartlett (\citeyear{Bar57}) and Jeffreys
(\citeyear{Jef61}). As Dawid (\citeyear{Daw11}) pointed out, the Bayes
factor is only one of the two elements on the right side of (\ref{bf})
which contribute toward the posterior model probabilities. The prior
model probabilities are of equal significance. By focusing on the
impact of the prior distributions for model parameters on the Bayes
factor, there is an implicit understanding that the prior model
probabilities are specified independently of these prior distributions.
This is \mbox{often} the case in practice, where a uniform prior distribution
over models is commonly adopted, as a reference position. Examples
where nonuniform prior distributions have been suggested include the
works of Madigan et al. (\citeyear{autokey37}), Chipman
(\citeyear{Chi96}), Laud and Ibrahim (\citeyear{LauIbr95,LauIbr96}),
Chipman, George and McCulloch (\citeyear{ChiGeoMcC01}), Cui and George
(\citeyear{CuiGeo08}), Ley and Steel (\citeyear{LeySte09}) and Wilson
et al. (\citeyear{Wiletal10}). We propose a different approach where we
consider how the two elements of the prior distribution under model
uncertainty might be jointly specified so that perceived problems with
Bayesian model comparison can be avoided. This leads to a nonuniform
specification for the prior distribution over models, depending
directly on the prior distributions for model parameters.

A related issue concerns the use of improper prior distributions for
model parameters. Such prior distributions involve unspecified
constants of proportionality, which do not appear in posterior
distributions for model parameters but do appear in the marginal
likelihood for any model and in any associated Bayes factors, so these
quantities are not uniquely determined. There have been several
attempts to address this issue, and to define an appropriate Bayes
factor for comparing models with improper priors; see Kadane and Lazar
(\citeyear{KadLaz04}) for a review. In such examples, Dawid
(\citeyear{Daw11}) proposed that the product of the prior model
``probability'' and the prior density for a given model could be
determined simultaneously by eliciting the relative prior
``probabilities'' of particular sets of parameter values for different
models. He also suggested an approach for constructing a general
noninformative prior, over both models and model parameters, based on
Jeffreys priors for individual models. Although the prior distributions
for individual models are not generally proper, they have densities
which are uniquely determined and hence the posterior distribution over
models can be evaluated. Clyde (\citeyear{Cly00}) proposed a similar
approach where the priors for parameters of individual models are
uniform and the relative weights of different models are chosen by
constraining the resulting posterior model probabilities to be
equivalent to those resulting from a~specified information criterion,
such as BIC.

Here, we do not consider improper prior distributions for the model
parameters, but our approach is similar in spirit as we do explicitly
consider a joint specification of the prior over models and model
parameters.

We focus on models in which the parameters are sufficiently homogeneous
(perhaps under transformation) so that a multivariate normal prior
density $N(\bolds{\mu}_m, V_m)$ is appropriate, and in which the
likelihood is sufficiently regular for standard asymptotic results to
apply. Examples are linear regression models, generalized linear models
and standard time series models. In much of what follows, with
minor~modification, the normal prior can be replaced by any elliptically
symmetric prior density proportional to
$|V|^{-1/2}g((\bb-\bolds{\mu})^TV^{-1}(\bb-\bolds{\mu}))$ where\break
$\int_0^\infty r^{d-1} g(r^2)\,dr<\infty$ and $d$ is the dimensionality
of~$\bb$. This includes prior distributions from the multivariate $t$
or Laplace families. Similarly, our approach can also be adapted to
common prior distributions for parameters of graphical models.

We choose to decompose the prior variance matrix as $V_m = c_m^2
\Sigma_m$ where $c_m$ represents the scale of the prior dispersion and
$\Sigma_m$ is a matrix with a~specified value of $|\Sigma_m|$,
although for the remainder of this section we do not require an
explicit value; further discussion of this issue is presented in
Section~\ref{sec2}. Hence, suppose that
%
\begin{eqnarray}\label{normal_prior}
&&f(\bb_m|m)\nonumber\hspace*{-20pt}
\\
&&\quad=(2\pi)^{-d_m/2} |\Sigma_m|^{-1/2} c_m^{-d_m}\hspace*{-20pt}
\\
&&\qquad{}\cdot\exp\biggl(-\frac{1}{2c_m^2}(\bb_m-\bolds{\mu}_m)^T\Sigma_m^{-1}(\bb
_m-\bolds{\mu}_m)\biggr).\nonumber\hspace*{-20pt}
\end{eqnarray}
Then,
%
\begin{eqnarray}\label{general}
f(m|\mathbf{y})&\propto&f(m)\int f(\mathbf{y} |m, \bolds{\beta}_m)
f(\bolds{\beta}_m|m)\,d\bolds{\beta}_m\nonumber\\
&=& f(m)(2\pi)^{-d_m/2} |\Sigma_m|^{-1/2} c_m^{-d_m}\nonumber
\\[-8pt]\\[-8pt]
&&{}\cdot
\int_{\rr^{d_m}}\exp\biggl(-\frac{1}{2c_m^2}(\bb_m-\bolds{\mu}_m)^T\Sigma
_m^{-1}\nonumber
\\
&&{}\hphantom{\int_{\rr^{d_m}}\exp\biggl(}\cdot(\bb_m-\bolds{\mu}_m)\biggr) f(\mathbf{y}|m,
\bolds{\beta}_m)\,d\bolds{\beta}_m\nonumber
\end{eqnarray}
and for suitably large $c_m$,
%
\begin{eqnarray}\label{general2}
f(m|\mathbf{y}) &\approx& f(m)
(2\pi)^{-d_m/2} |\Sigma_m|^{-1/2} c_m^{-d_m}\nonumber
\\[-8pt]\\[-8pt]
&&{}\cdot\int_{\rr^{d_m}}
f(\mathbf{y} |m, \bolds{\beta}_m)\,d\bolds{\beta}_m.\nonumber
\end{eqnarray}
Hence, as $c_m^2$ gets larger, $f(m|\by)$ gets smaller, assuming
everything else remains fixed. Therefore, for two models of different
dimension with the same value of $c_m^2$, the posterior odds in favor
of the more complex model tend to zero as $c_m^2$ gets larger, that is,
as the prior dispersion increases at a common rate. This is essentially
Lindley's paradox.

There have been substantial recent computational advances in
methodology for exploring the model space; see, for example, Green
(\citeyear{Gre95,Gre03}), Kohn, Smith and Chan (\citeyear{KohSmiCha01}), Denison et al.
(\citeyear{Denetal02}), Hans, Dobra and West
(\citeyear{HanDobWes07}). The related discussion of the important problem of choosing
prior parameter dispersions has been largely focused on ways to avoid
Lindley's paradox; see, for example, Fern{\'a}ndez, Ley and
Steel (\citeyear{FerLeySte01}) and Liang
et al. (\citeyear{Liaetal08}) for detailed discussion on appropriate choices of
$g$-priors for linear regression models and Raftery (\citeyear{Raf96}) and
Dellaportas and Forster (\citeyear{DelFor99}) for some guidelines on selecting
dispersion parameters of normal priors for generalized linear model
parameters. Other approaches which have been proposed for specifying
default prior distributions under model uncertainty which provide
\mbox{plausible} posterior model probabilities include intrinsic priors
(Berger and Pericchi, \citeyear{BerPer96}) and, for normal linear models, mixtures of
$g$-priors (Liang et al., \citeyear{Liaetal08}). The important effect that any of these
prior specifications might have on the parameter posterior
distributions within each model has been largely neglected. For
example, a~set of values of $c_m$ might be appropriate for addressing
model uncertainty, but might produce prior densities $f(\bb_m|m)$ that
are insufficiently diffuse and overstate prior information within
certain models. This has a serious effect on posterior and predictive
densities of all\vadjust{\goodbreak} quantities of interest in any data analysis. This is a
particularly important consideration when posterior or predictive
inferences are integrated over models (model-averaging). In such
analyses both the prior model probabilities and prior distributions
over model parameters can have a significant impact on inferences.

In this paper we propose that prior distributions for model parameters
should be specified with the issue of inference conditional on a
particular model being the primary focus. For example, when only
weak information concerning the model parameters is available, a
highly diffuse prior may be deemed appropriate. The key element of our
proposed approach is that sensitivity of posterior model probabilities
to the exact scale of such a diffuse prior is avoided by suitable
specification of prior model probabilities $f(m)$. As mentioned above,
these probabilities are rarely specified carefully, a discrete uniform
prior distribution across models usually being adopted. However, it is
straightforward to see that setting $f(m) \propto c_m^{d_m}$ in
(\ref{general2}) will have the effect of eliminating dependence of the
posterior model probability $f(m|y)$ on the prior dispersion $c_m$.
This provides a motivation for investigating how prior model
probabilities can be chosen in conjunction with prior distributions for
model parameters, by first considering properties of the resulting
posterior distribution.

The strategy described in this paper can be viewed as a full Bayesian
approach where the prior distribution for model parameters is specified
by focusing on the uncertainty concerning those parameters alone, and
the prior model probabilities can be specified by considering the way
in which an associated ``information criterion'' balances parsimony and
goodness of fit. In the past, informative specifications for these
probabilities have largely been elicited via the notion of imaginary
data; see, for example, Chen, Ibrahim and
Yiannoutsos (\citeyear{CheIbrYia99}) Chen et al. (\citeyear{Cheetal03}). Within the approach
suggested here, prior model probabilities are specified by considering
the way in which data yet to be observed might modify one's beliefs
about models, given the prior distributions for the model parameters.
Full posterior inference under model uncertainty, including model
averaging, is then available for the chosen prior.

\section{Prior and Posterior Distributions} \label{sec2}

We consider the joint specification of the two components of the prior
distribution by investigating its impact on the asymptotic posterior
model probabilities. This allows us to investigate, across a wide class
of models, the sensitivity of posterior inferences to the specification
of prior model probabilities and prior distributions for model
parameters.
By using Laplace's method to approximate the
posterior marginal likelihood in (\ref{general}), we obtain, subject to
certain regularity conditions (see Kass, Tierney and
Kadane, \citeyear{KasTieKad88}; Schervish, \citeyear{Sch95},
Section~7.4.3)
%
\begin{eqnarray}
f(m|\by)&\propto& f(m) |\Sigma_m|^{-1/2} c_m^{-d_m} f(\mathbf
{y}|m,\widehat{\bolds{\beta}}_m)\nonumber
\\
&&{}\cdot\exp\biggl(-\frac
{1}{2c_m^2}(\widehat{\bb}_m-\bolds{\mu}_m)^T\Sigma_m^{-1}(\widehat{\bb
}_m-\bolds{\mu}_m)\biggr) \\
&&{}\cdot |c_m^{-2} \Sigma_m^{-1} -H(\widehat{\bb}_m)|^{-1/2} \bigl( 1+
O_{p}(n^{-1}) \bigr),\nonumber
\end{eqnarray}
where $n$ is the sample size, $\widehat{\bb}_m$ is the maximum
likelihood estimate and $H(\bb_m)$ is the second derivative matrix for
$\log f(\mathbf{y}|m, \bolds{\beta}_m)$. Then,
%
\begin{eqnarray}\label{logpm}
&&\log f(m|\by)\nonumber
\\[-0.5pt]
&&\quad= C+ \log f(m) - \frac{1}{2} \log|\Sigma_m| - d_m \log c_m\nonumber
\\[-0.5pt]
&&\qquad{}+ \log
f(\mathbf{y}|m, \widehat{\bolds{\beta}}_m)\nonumber\\[-0.5pt]
&&\qquad{} -\frac{1}{2c_m^2}(\widehat{\bb}_m-\bolds{\mu}_m)^T\Sigma
_m^{-1}(\widehat{\bb}_m-\bolds{\mu}_m)\nonumber
\\[-0.5pt]
&&\qquad{}- \frac{1}{2} \log|c_m^{-2}
\Sigma
_m^{-1}-H(\widehat{\bb}_m)| + O_{p}(n^{-1}) \\[-0.5pt]
&&\quad= C+ \log f(m)\nonumber
\\[-0.5pt]
&&\qquad{} - \frac{1}{2} \log|\Sigma_m| - d_m \log c_m+ \log
f(\mathbf{y}|m, \widehat{\bolds{\beta}}_m)\nonumber \\[-0.5pt]
&&\qquad{}
-\frac{1}{2c_m^2}(\widehat{\bb}_m-\bolds{\mu}_m)^T\Sigma_m^{-1}(\widehat
{\bb}_m-\bolds{\mu}_m)-\frac{d_m}{2} \log n \nonumber
\\[-0.5pt]
&&\qquad{}- \frac{1}{2}
\log|i(\widehat{\bb}_m)| + O_{p}(n^{-1/2}),\nonumber
\end{eqnarray}
where $C$ is a normalizing constant to ensure that the posterior model
probabilities sum to 1 and $i(\bb_m) \approx- n^{-1} H(\bb_m)$ is the
Fisher information matrix for a~unit observation; see Kass and
Wasserman (\citeyear{KasWas95}).

We propose specifying the decomposition of the prior variance matrix
$c^2_m \Sigma_m$ so that $|\Sigma_m|\!=\!|i(\bb_m)|^{-1}$, resulting in
%
\begin{eqnarray}\label{logpm1}
\log f(m|\by) &=& C + \log f(\mathbf{y}|m,
\widehat{\bolds{\beta}}_m)\nonumber
\\
&&{}-\frac{1}{2c_m^2}(\widehat{\bb}_m-\bolds{\mu}
_m)^T\Sigma_m^{-1}(\widehat{\bb}_m-\bolds{\mu}_m)\nonumber
\\[-8pt]\\[-8pt]
&&{} + \log
f(m) - d_m \log c_m\nonumber
\\
&&{}- \frac{d_m}{2} \log n + O_{p}(n^{-1/2}),\nonumber
\end{eqnarray}
where $c_m^{-2}$ defined as
%
\begin{equation}\label{unitpr} c_m^{-2}=(
|V_m||i(\bb_m)|)^{-1/d_m}\vadjust{\goodbreak}
\end{equation}
can be interpreted as the number of units of information in the
prior.

Note that substituting $c_m=1$ (unit information) into (\ref{logpm1}),
and choosing a discrete uniform prior distribution across models,
suggests model comparison on the basis of a modified version of the
Schwarz criterion (BIC; Schwarz, \citeyear{Sch78}) where maximum likelihood is
replaced by maximum penalized likelihood. In a comparison of two nested
models, Kass and Wasserman (\citeyear{KasWas95}) gave extra conditions on a unit
information prior which lead to model comparison asymptotically based
on BIC; see Volinsky and Raf\-tery (\citeyear{VolRaf00}) for an example of the use of
unit information priors for Bayesian model comparison. For
regression-type models where the components of $\by$ are not
identically distributed, depending on explanatory data, the unit
information as defined above potentially changes as the sample size
changes, so a~little care is required with asymptotic arguments. We
assume that the explanatory variables arise in such a way that
$i(\bb_m)=i_{\lim}(\bb_m)+O(n^{-1/2})$ where $i_{\lim}(\bb_m)$ is
a finite limit. This is not a great restriction and is true, for
example, where the explanatory data may be thought of as
i.i.d.~observations from a~distribution with finite variance.

In general, $i(\bb_m)$ depends on the unknown model parameters, so the
number of units of information~$c_m^{-2}$ corresponding to any given
prior variance matrix $V_m$ will also not be known, and hence it is not
generally possible to construct an exact unit information prior.
Dellaportas and Forster (\citeyear{DelFor99}) and Ntzoufras, Dellaportas and
Forster (\citeyear{NtzDelFor03}) advocated
substituting $\bolds{\mu}_m$, the prior mean of $\bb_m$, into
$i(\bb_m)$ to give a~prior for model comparison which has a unit
information interpretation but for which model comparison is not
asymptotically based on BIC.

When the prior distribution for the parameters of model $m$ is highly
diffuse, so that $c_m$ is large, then~(\ref{logpm1}) can be rewritten
as
%
\begin{eqnarray}\label{logpm2}
\log f(m|\by) &\approx& C + \log f(\mathbf{y}|m,
\widehat{\bolds{\beta}}_m) \nonumber
\\[-3pt]\\[-12pt]
&&{}+ \log f(m) - d_m \log c_m- \frac{d_m}{2}
\log n,\nonumber\hspace*{-20pt}
\end{eqnarray}
where $\widehat{\bb}_m$ is the maximum
likelihood estimate of~${\bb}_m$. Equation (\ref{logpm2}) corresponds
asymptotically to an information criterion with complexity penalty
equal to $\log n+\log c_m^2 - 2 d_m^{-1} \log f(m)$ compared with BIC,
for example, where the complexity penalty is equal to $\log n$. The
relative discrepancy between these two penalties is asymptotically
zero. Poskitt and Tre\-mayne (\citeyear{PosTre83}) discussed the
interplay between prior
model probabilities\vadjust{\goodbreak} $f(m)$ and BIC and other information criteria in a
time series context when Jeffreys priors are used for model parameters.

It is clear from (\ref{logpm2}) that a large value of $c_m$ arising
from a diffuse prior penalizes more complex models. On the other hand,
a more moderate value of $c_m$ (such as unit information) may have the
effect of shrinking the posterior distributions of the model parameters
toward the prior mean to a greater extent than desired. This has a
particular impact when model averaging is used to provide predictive
inferences (see, e.g., Hoeting et al., \citeyear{Hoeetal99}), where both the
posterior model probabilities and the posterior distributions of the
model parameters are important. A conflict can arise where to achieve
the amount of dispersion desired in the prior distribution for model
parameters, more complex models are unfairly penalized. To avoid this,
we suggest choosing the dispersion of the prior distributions of model
parameters to provide the amount of shrinkage to the prior mean which
is considered appropriate a priori, and to choose prior model
probabilities to adjust for the resulting effect this will have on the
posterior model probabilities. We propose
%
\begin{equation}\label{pmp}f(m) \propto p(m)
c_m^{d_m},
\end{equation}
where $p(m)$ are baseline model
probabilities. The~purpose of decomposing prior model
probabilities~$f(m)$ in this way is to explicitly specify a direct dependence between
these probabilities and the hyperparameters of the prior distributions
for the parameters of each model. There is no requirement that~$p(m)$
be uniform, and any differences between~$f(m)$ for different~$m$ which
are unrelated to the prior distributions for the model parameters are
absorbed in~$p(m)$. Often, we might expect~$p(m)$ not to depend on the
dimensionalities of the models, although we do not prohibit this. With
this choice of $f(m)$, (\ref{logpm1}) becomes
%
\begin{eqnarray}\label{logpmp2}
\log f(m|\by) &=& C + \log f(\mathbf{y}|m, \widehat{\bolds{\beta
}}_m)\nonumber
\\
&&{}-\frac{1}{2c_m^2}(\widehat{\bb}_m-\bolds{\mu}_m)^T\Sigma
_m^{-1}(\widehat{\bb}_m-\bolds{\mu}_m)
\\
&& + \log p(m) - \frac{d_m}{2} \log n + O_{p}(n^{-1/2}),\nonumber
\end{eqnarray}
where the specification of the base variance $\Sigma_m$ is not in terms
of unit information, the extra term $-\log(|\Sigma_m|\cdot|i(\bb_m)|)/2$ is
required in (\ref{logpmp2}). When $c_m^2$ is large and when all $p(m)$
are equal, model comparison is asymptotically based on BIC. More
generally, we propose choosing prior model probabilities based on
(\ref{pmp}) for any\vadjust{\goodbreak} prior variance $V_m$. Substituting (\ref{unitpr})
into (\ref{pmp}), we obtain
%
\begin{equation}\label{pmp2}
f(m)\propto p(m) (|V_m|
|i(\bb_m)|)^{1/2}.
\end{equation}
The choice of $p(m)$ can be based on the form of the equivalent model
complexity penalty which is deemed to be appropriate a priori. Setting
all $p(m)$ equal, which we propose as the default option, leads to
model determination based on a modified BIC criterion involving
penalized maximum likelihood. Hence, the impact of the prior
distribution on the posterior model probability through
$(\widehat{\bb}_m-\break\bolds{\mu}_m)^T\Sigma_m^{-1}(\widehat{\bb}_m-\bolds{\mu}_m)/2c_m^2$
in (\ref{logpmp2}) is straightforward to assess, and any undesirable
side effects of large prior variances are eliminated. In
Section~\ref{sec1}, we discussed existing approaches for specifying
nonuniform $f(m)$ based on considerations such as the desire to control
model size. These can easily be incorporated into the specification of
nonuniform $p(m)$, if desired. Other possible approaches to specifying
or eliciting $p(m)$ are discussed in
Sections~\ref{sec4}~and~\ref{sec5}.

In order to specify prior model probabilities using~(\ref{pmp}), with
$p(m)$ chosen to correspond to a particular complexity penalty, it is
necessary to be able to evaluate $c_m^{-2}$, the number of units of
information implied by the specified prior variance~$V_m$ for~$\bb_m$.
Equivalently, as $f(m)\propto p(m) |V_m|^{1/2}
|i(\bb_m)|^{1/2}$,\break knowledge of $|i(\bb_m)|$ is required.
Except in certain circumstances, such as normal linear models, this
quantity depends on the unknown model parameters $\bb_m$. This is not
appropriate as a specification for the marginal prior distribution over
model space. One possibility is to use a sample-based estimate
$|i(\widehat{\bb}_m)|$ to determine the ``prior'' model probability, in
which case the approach is not fully Bayesian. Alternatively, as
suggested above, substituting $\bolds{\mu}_m$, the prior mean of
$\bb_m$, into $i(\bb_m)$ gives a prior for model comparison which has a
unit information interpretation but for which model comparison is not
asymptotically based on (\ref{logpmp2}), the extra term
$\log(|i(\bolds{\mu}_m)|/\allowbreak |i(\bb_m)|)/2$ being required.

\section{Normal Linear Models} \label{normal_models}

Here we consider normal linear models where for $m\in M$, $\by\sim
N(\mathbf{X}_m\bb_m, \sigma^2 I)$ with the conjugate prior
specification
%
\begin{eqnarray}\label{nig}
\bb_m|\sigma^2, m&\sim& N(\bolds{\mu}_m, \sigma^2V_m)
\quad\mbox{and}\nonumber
\\[-8pt]\\[-8pt]
 \sigma^{-2}&\sim&\operatorname{Gamma}(\alpha, \lambda).\nonumber
\end{eqnarray}
For such models the posterior model probabilities can be calculated\vadjust{\goodbreak}
exactly. Dropping the model subscript $m$ for clarity,
\begin{eqnarray*}
&&f(m|\by)
\\
&&\quad\propto f(m)
\frac{|V^*|^{1/2}}{|V|^{1/2}}
\\
&&\qquad{}\cdot\bigl(2\lambda+\by^T\by+\bolds{\mu}^T V^{-1}
\bolds{\mu}-
\widetilde{\bb}^T(V^*)^{-1}\widetilde{\bb}\bigr)^{-\alpha-n/2},
\end{eqnarray*}
where $V^* = ( V^{-1} + \bX^T \bX)^{-1}$ and $\widetilde{\bb} = V^*
(V^{-1} \bolds{\mu}+ \bX^T \by)$ is the posterior mean. Hence, setting
$V=c^2\Sigma$, as before,
%
\begin{eqnarray}
\label{mlexact}
&&\log f(m|\by)\nonumber\hspace*{-15pt}\\
&&\quad =  C + \log f(m) -\frac{1}{2} \log| c^{-2} \Sigma
^{-1} +\bX^{T}\bX|\nonumber\hspace*{-15pt}
\\
&&\qquad{} -\frac{1}{2} \log|\Sigma|-d \log c\nonumber\hspace*{-15pt}\\
& &\qquad{} - ( \alpha+n/2 ) \log\bigl(2\lambda+\by^T\by+\bolds{\mu} ^T V^{-1}
\bolds{\mu}\hspace*{-15pt}
\\
&&\qquad\hphantom{- ( \alpha+n/2 ) \log\bigl(}{}-\widetilde{\bb}^T(V^*)^{-1}\widetilde{\bb}\bigr)
\nonumber\hspace*{-15pt}
 \\
\label{normal}&&\quad =  C - ( \alpha+n/2 ) \log\bigl(2\lambda+(\by-\bX
\widetilde{\bb})^T(\by-\bX\widetilde{\bb})\nonumber\hspace*{-15pt}
\\
&&\qquad\hphantom{C - ( \alpha+n/2 ) \log\bigl(}{}+(\widetilde{\bb}-\bolds{\mu}
)^T V^{-1}(\widetilde{\bb}-\bolds{\mu}) \bigr) \nonumber\\[-8pt]\\[-8pt]
&&\qquad{} +\log f(m) -\frac{1}{2} \log| i | \nonumber\hspace*{-15pt}
\\
&&\qquad{}- \frac{d}{2} \log n -\frac{1}{2}
\log|\Sigma|-d \log c + O(n^{-1}),\nonumber\hspace*{-15pt}
\end{eqnarray}
where, with a slight abuse of notation, $i=n^{-1}\bX^T \bX$ is the
unit information matrix multiplied by $\sigma^2$. Notice the
correspondence between (\ref{logpm}) and (\ref{normal}). As before, if
$|\Sigma|=|i|^{-1}$, then $c^{-2}$ can be interpreted as the number of
units of information in the prior (as the prior variance is $c^2
\sigma^2 \Sigma$) and
%
\begin{eqnarray}\label{normal2}
&&\log f(m|\by)\nonumber\hspace*{-15pt}
\\
 &&\quad =  C - ( \alpha+n/2 ) \log
\bigl(2\lambda+(\by-\bX\widetilde{\bb})^T(\by-\bX\widetilde{\bb
})\nonumber\hspace*{-15pt}
\\[-8pt]\\[-8pt]
&&\qquad\hphantom{C - ( \alpha+n/2 ) \log
(}{}+(\widetilde{\bb}-\bolds{\mu})^T V^{-1}(\widetilde{\bb}-\bolds{\mu})\nonumber
\bigr)\hspace*{-15pt}
\\
&&\qquad{}+ \log f(m) - \frac{d}{2} \log n -d \log c + O(n^{-1}).\nonumber\hspace*{-15pt}
\end{eqnarray}
In both (\ref{normal}) and (\ref{normal2}) the posterior mean
$\widetilde{\bb}$ can be replaced by the least squares estimator
$\widehat{\bb}$. Again, if $c=1$ (unit information) and the prior
distribution across models is uniform, model comparison is performed
using a modified version of BIC, as presented for example by Raftery
(\citeyear{Raf95}), where $n/2$ times the logarithm of the residual sum of squares
for the model has been replaced by the first term on the right-hand side of
(\ref{normal2}). The residual sum of squares is evaluated at the\vadjust{\goodbreak}
posterior mode, and is penalized by a term representing deviation from
the prior mean, as in (\ref{logpm}). This expression also depends on
the prior for $\sigma^2$ through the prior parameters~$\alpha$
and~$\lambda$, although these terms vanish when the improper prior
$f(\sigma^2)\propto\sigma^{-2}$, for which $\alpha=\lambda=0$, is
used. With these values, and setting $\Sigma^{-1}=i=n^{-1}\bX^T\bX$,
we obtain the prior used by Fern{\'a}ndez, Ley and
Steel (\citeyear{FerLeySte01}), who also noted the
unit information interpretation when $c=1$ for all $m$. This is an
example of a $g$-prior (Zellner, \citeyear{Zel86}).

As before, if the prior variance $V$ suggests a different value of
$c$, then the resulting impact on the posterior model probabilities can
be moderated by an appropriate choice of $f(m)$ and again we propose
the use of (\ref{pmp}) and (\ref{pmp2}), noting that for normal models
$i$ is known. In the context of normal linear models, Pericchi (\citeyear{Per84})
suggested a similar adjustment of prior model probabilities by an
amount related to the expected gain in information. Alternatively,
replacing $|i|$ by $|i+n^{-1}V^{-1}|$ in (\ref{pmp2}), resulting in
%
\begin{equation}\label{pmp3}
f(m)\propto p(m) |V|^{1/2} |i+n^{-1}V^{-1}|^{1/2},
\end{equation}
makes (\ref{normal}) exact, eliminating the
$O(n^{-1})$ term. Again, for highly diffuse prior distributions on the
model parameters (large values of $c^2$), together with
$\alpha=\lambda=0$ and prior model probabilities based on (\ref{pmp})
and (\ref{pmp2}), equation (\ref{normal2}) implies that model
comparison is performed on the basis of BIC.

We note that when the $g$-prior $\Sigma^{-1}=i=n^{-1}\bX^T\bX$ is used,
together with $\bolds{\mu}={\bf0}$, then the posterior model
probability (\ref{mlexact}) can be written as
%
\begin{eqnarray}\label{ours}
&&\log f(m|\by)\nonumber
\\
&&\quad=C + \log f(m) -\frac{d}{2} \log(n+c^{-2})-d \log
c\nonumber\\[-8pt]\\[-8pt]
& &\qquad{}- ( \alpha+n/2 ) \log \biggl(2\lambda+\frac{1}{1+nc^2}\by^T\by\nonumber
\\
&&\qquad\hphantom{- ( \alpha+n/2 ) \log \biggl(}{}+\frac
{nc^2}{1+nc^2}S^2_y(1-R^2)\biggr),\nonumber
\end{eqnarray}
where $S^2_y=\sum_{i=1}^n (y_i-\bar y)^2$ and $R^2$ is the standard
coefficient of determination for the model. For our prior, where
$f(m)\propto p(m)c^d$, we obtain
\begin{eqnarray*}
&&\log f(m|\by)
\\
&&\quad=C + \log p(m) -\frac{d}{2} \log(n+c^{-2})
\\
&&\qquad{}- ( \alpha+n/2
) \log \biggl(2\lambda+\frac{1}{1+nc^2}\by^T\by
\\
&&\qquad\hphantom{- ( \alpha+n/2
) \log \biggl(}{}+\frac
{nc^2}{1+nc^2}S^2_y(1-R^2)\biggr).
\end{eqnarray*}
The trade-off between model fit, as reflected by $R^2$, and complexity,
measured by $d$, is immediately apparent, with the complexity penalty
tending to BIC as $c^{-2}$ tends to zero. The posterior model
probability (\ref{ours}) is similar to expression (5) of Liang et al.
(\citeyear{Liaetal08}). Their approach differs in that they consider the intercept
parameter of the linear model separately, giving it an improper uniform
prior, as this parameter is common to all models under consideration.
Such a specification might also be adopted within our framework, both
for linear models and for more general regression models.

\section{Specification of $\lowercase{p(m)}$ Based on Relationship
with Other Information Criteria}\label{sec4}

In Sections \ref{sec2} and \ref{normal_models}, we have investigated
how prior model probabilities might be specified by considering their
joint impact, together with the prior distributions for the model
parameters, on the posterior model probabilities. It was shown that
making these probabilities depend on the prior variance of the
associated model parameters using (\ref{pmp}) or (\ref{pmp2}) with
uniform $p(m)$ leads to posterior model probabilities which are
asymptotically equivalent (to order~$n^{-1/2}$) to those
implied by BIC. For models other than normal linear regression models,
a prior value of $\bb$ must be substituted into (\ref{pmp2}) and so the
approximation only attains this accuracy for $\bb$ within an
$O(n^{-1/2})$ neighborhood of this value. Nevertheless, we
might expect BIC to more accurately reflect the full Bayesian analysis
for such a prior than more generally, where the error of BIC as an
approximation to the log-Bayes factor is $O(1)$.

Alternative (nonuniform) specifications for $p(m)$ might be based on
matching the posterior model probabilities (\ref{logpm1}) using prior
weights (\ref{pmp2}) with other information criteria of the form
\[
\log f(\by|m, \widehat{\bb}_m) - \tfrac{1}{2} \psi(n) d_m,
\]
where $\psi(n)$ is a ``penalty'' function; for BIC, $\psi(n)=\log n$
and for AIC $\psi(n)=2$. From (\ref{logpmp2}), for large $c_m^2$ or for
a modified criterion, we have $\psi(n)=\log n+2d_m^{-1}\log p(m)$. As
$p(m)$ contributes to the prior model probability through (\ref{pmp})
it cannot be a function of~$n$ since our prior belief on models should
not change as the sample size changes. Therefore, strictly, the only
penalty functions which can be equivalent to setting prior model
probabilities as in (\ref{pmp}) are of the form $\psi(n)=\log n+\psi_0$
for some positive constant \mbox{$\psi_0>0$}. Any alternative dependence on
$n$ would correspond to a prior which\vadjust{\goodbreak} depended on $n$, through~$f(m)$
or $f(\bb_m|m)$. Hence AIC, for example, is prohibited (as would be
expected since AIC is not consistent), whereas any approach arising
from a proper prior must be consistent. Nevertheless, if a penalty
function of a~particular form is desired for a sample of a~specified
size $n_0$, then setting $\log p(m)=\frac{d_m}{2} \{ \log n_0 -
\psi(n_0) \}$ will ensure that posterior model probabilities are
calculated on the basis of the information criterion with penalty
$\psi(n_0)$, at the relevant sample size~$n_0$.

Clyde (\citeyear{Cly00}) proposed CIC, a calibrated information criterion, based on
a joint specification of (improper) uniform prior distributions for
model parameters, together with prior model probabilities
\[
f(\bb_m|m)f(m)\propto(2\pi)^{-d_m/2}\biggl|\frac{n}{c}
i(\widehat{\bb}_m)\biggr|^{1/2},
\]
where $c$ is a constant which is determined by constraining the
posterior model probabilities to be the same as those which would arise
from an alternative information criterion, such as BIC. For our prior,
in the limit as $c_m^{-2}\to0$, we have
\[
f(\bb_m|m)f(m)\propto(2\pi)^{-d_m/2}|\Sigma_m|^{-1/2}p(m)
\]
so in the case where $|\Sigma_m|=|i(\bb_m)|$ for a value of $\bb_m$
close to the m.l.e.~these approaches will yield similar results if
$p(m)$ is calibrated to $(n/c)^{d/2}$, which is plausible if $c\propto
n$. Note also that, if $p(m)\propto\break (2\pi)^{d_m/2}|\Sigma_m|^{1/2}$,
our prior in this limiting case reduces to a uniform measure over the
``parameter space'' for $(m,\bb_m)$.

\section{\texorpdfstring{Alternative Arguments for $\lowercase{f(m)\propto c_m^{d_m}}$}
{Alternative Arguments for f(m) propto c_m^{d_m}}} \label{sec5}

The purpose of the following discussion is not to advocate a particular
prior, but simply to illustrate that one can arrive at (\ref{pmp}) by
direct consideration of prior probabilities, or prior densities, or by
the behavior of posterior means, as well as by the asymptotic behavior
of posterior model probabilities, or associated numerical
approximations, as earlier.

\subsection{Constant Probability in a Neighborhood of the Prior Mean}

Specifying the prior distribution on the basis of how it is likely to
impact the posterior distribution is entirely valid, but may perhaps
seem unnatural. In particular, the consequence that the prior model
probabilities might depend on the prior distributions for the model
parameters may seem somewhat alien. This is particularly true of the
implication of (\ref{pmp2}),\vadjust{\goodbreak} that models where we have more
information (smaller dispersion) in the prior distribution should be
given lower prior probabilities than models for which we are less
certain about the parameter values. One justification for this is to
examine the prior model probabilities for particular subsets of the
parameter spaces within models. This can be considered as an extension
of the approach of Robert (\citeyear{Rob93}) for two normal models. We consider
the prior probability of the event
\begin{eqnarray*}
E&=&\{\mbox{model } m \mbox{ is `true'} \}
\\
&&{}\cap
\{(\bb_m-\bolds{\mu}_m)^Ti(\bb^0_m)(\bb_m-\bolds{\mu}_m)<\varepsilon^2\}
\end{eqnarray*}
for some reference parameter value $\bb^0_m$, possibly the prior mean
$\bolds{\mu}$. The dependence of this subset of the parameter space on
the unit information at $\bb^0_m$ enforces some degree of comparability
across models. This is particularly true if the various values of
$\bb^0_m$ are compatible (e.g., they imply the same linear predictor in
a generalized linear model, as they would generally do if set equal to
$\bf0$). For the purposes of the current discussion, we also require
$V_m=c_m^2 i(\bb^0_m)^{-1}$. This is a plausible default choice, but
nevertheless represents considerable restriction on the structure of
the prior variance, which was previously unconstrained. Then
\begin{eqnarray*}
P(E) &=&f(m)P \biggl(\chi^2_{d_m}< \frac{ \varepsilon^2 }{ c_m^2 } \biggr)
\\
&\approx&\frac{f(m)\varepsilon^{d_m}}{2^{d_m/2-1}\Gamma(d_m/2)c_m^{d_m}}
\end{eqnarray*}
for small $\varepsilon$. Therefore, for this prior, if the joint prior
probability of model $m$ in conjunction with~$\bb_m$ being in some
specified neighborhood (defined according to a unit information inner
product) of its prior mean is to be uniform across models, then we
require $f(m)\propto p(m) c_m^{d_m}$ as in (\ref{pmp}), with $p(m)=
2^{d_m/2-1}\Gamma(d_m/2)/\varepsilon^{d_m}$.

\subsection{Flattening Prior Densities}

An alternative justification of (\ref{pmp}) when the model parameters
are given diffuse normal prior distributions arises as follows. One way
of taking a ``baseline'' prior distribution and making it more diffuse,
to represent greater prior uncertainty, is to raise the prior density
to the power $1/c^2$ for some $c^2>1$, and then renormalize. For
example, for a single normal distribution this has the effect of
multiplying the variance by $c^2$, which increases the prior dispersion
in an obvious way. Highly diffuse priors, suitable in the absence of
strong prior information, may be thought of as arising from a baseline
prior transformed in this way for some large\vadjust{\goodbreak} value of $c^2$. Where
model uncertainty exists, the joint prior distribution is a mixture
whose components correspond to the models, with mixture weights $f(m)$.
As suggested above, a diffuse prior distribution might be obtained by
raising a baseline prior density (with respect to the natural measure
over models and associated parameter spaces) to the power $1/c^2$ and
renormalizing. Where the baseline prior distribution for $\bb_m$ is
normal with mean $\bolds{\mu}_m$ and variance $\bsm$, the effect of
raising the mixture prior density to the power $1/c^2$ is to increase
the variance of each~$\bb_m$ by a factor of~$c^2$, as before. For large
values of $c^2$ the effect of the subsequent renormalization is that
the model probabilities are proportional to
$|\bsm|^{1/2}(2\pi)^{d_m/2}c^{d_m}$, independent of the model
probabilities in the original baseline mixture prior. Again this
illustrates a relationship between prior model probabilities and prior
dispersion parameters satisfying (\ref{pmp}). For the two normal models
considered by Robert (\citeyear{Rob93}) the resulting prior model probabilities are
identical. Where the baseline variance is based on unit information,~so
$|\bsm|=|i(\bb_m)|$, then the prior model probabilities~can be written
as (\ref{pmp2}) with $p(m)=(2\pi)^{d_m/2}|i(\bb_m)|^{-1/2}$.

\subsection{Bayesian Model Averaging and Shrinkage}

Finally, this approach can be justified by considering the behavior of
the posterior mean under model averaging. We restrict consideration
here to two nested models, $m_0$ and $m_1$, differing by a single
parameter $\beta$ and suppose that $f(y|m_0)=f(y|m_1,\beta_0)$. We
assume that the prior for $\beta$ under $m_1$ is $N(\beta_0, c^2 )$, so
the prior mean under model $m_1$ is the specified value of $\beta$
under model $m_0$, and, without loss of generality, we take
$\beta_0=0$. Under model $m_1$ the Bayes estimator for $\beta$ is the
posterior mean $E_1(\beta|y)$, which has asymptotic expansion
%
\begin{equation}\label{e1}
E_1(\beta|y)=\widehat\beta\biggl(1-\frac{i(\widehat\beta)}{n
c^2}\biggr)+\frac{a_3}{2i(\widehat\beta)^2n}+o(n^{-1}),\hspace*{-30pt}
\end{equation}
where $na_3$
is the third derivative of the log-likelihood, evaluated at
$\widehat\beta$ (see, e.g., Johnson, \citeyear{Joh70}; Ghosh, \citeyear{Gho94}). This
illustrates the usual effect of prior variance~$c^2$ and the
corresponding prior precision $c^{-2}$ as a~shrinkage parameter, with
the posterior mean being shrunk away from the m.l.e., with the amount
of shrinkage diminishing as $c^{-2}\to0$. Hence, for fixed $y$, the
posterior mean for $\beta$ is (asymptotically) monotonic in~$c^{-2}$.
Allowing for model uncertainty, we have\break
$E(\beta|y)=f(m_1|y)E_1(\beta|y)$ where
%
\begin{equation}\label{e2}
f(m_1|y)=\frac{1}{1+k(2\pi)^{1/2}cf_1(0|y)},
\end{equation}
where $f_1(\beta|y)$ is
the posterior (marginal) density for~$\beta$ under $m_1$, and $k$ are
the prior odds in favor of $m_0$ over $m_1$. Combining (\ref{e1}) and
(\ref{e2}), we see that, in $E(\beta|y)$, the model-averaged posterior mean
for~$\beta$, the m.l.e. $\hat{\beta}$ is multiplied by a shrinkage coefficient,
$f(m_1|y)E_1(\beta|y)$, which is not a monotonic function of the prior precision for
$\beta$ and hence $c^{-2}$ no longer has a simple interpretation as a shrinkage parameter.
A simple
\begin{figure}
\includegraphics{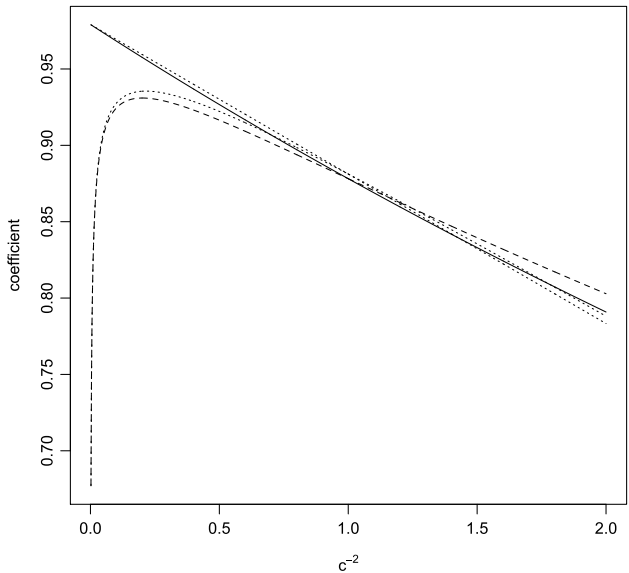}
\caption{Model average coefficient on $\widehat\beta$ [evaluated as
$\widehat\beta/\beta$], for normal likelihood with known error
variance, $\sigma^2$. The plot here is for $n=10,
\widehat\beta=1,\sigma^2=1$. The dashed line is for a uniform prior
over models, and the solid line uses prior model probability
$f(m_1)\propto c^{-1}$. The dotted lines are approximations based on
replacing $(2\pi)^{1/2}f_1(0|y)$ in (\protect\ref{e2}) with its normal
approximation $\exp(-\frac{i(\widehat\beta)n}{2}\widehat\beta^2)$,
ignoring the dependence, to $O(n^{-1})$, of $f_1(0|y)$ on $c^{-2}$.}
\label{figma}
\vspace*{-4pt}
\end{figure}
illustration of this is provided by Figure~\ref{figma}, where this
coefficient is plotted for various values of $c^{-2}$, for the simple
example of a~normal distribution with known error variance, and prior
odds $k=1$, corresponding to a uniform prior on model space. Note that
a high value of the coefficient on $\widehat\beta$ corresponds to low
shrinkage. It can be seen that, regardless of the value of $c^{-2}$,
there is a certain amount of shrinkage toward the prior mean and the
shrinkage is not a monotone function of $c^{-2}$. For values of
$c^{-2}$ greater than 0.5, the shrinkage to the prior mean is an
approximately linearly increasing function of $c^{-2}$ as expected. For
small values of $c^{-2}$, posterior probability is increasingly
concentrated on $m_0$ as $c^{-2}$ decreases (Lindley paradox) and hence
the model-averaged estimate is increasingly shrunk to the prior mean.
Adopting the approach advocated in this paper has the effect\vadjust{\goodbreak} of setting
$k\propto c^{-1}$ which mitigates this effect, and returns control over
the shrinkage to the analyst.

\section{Illustrated Examples} \label{sec-exam}

We illustrate our approach in a series of simulations and real data
applications. For comparison, we also present results under other
prior specifications, notably the hyper $g$-prior of Liang et al.
(\citeyear{Liaetal08}), for which computation is performed using the BAS package; see
Clyde (\citeyear{Cly10}).

Section \ref{ex2} illustrates that unit information prior
specifications (or other specifications suggesting\break smaller prior
parameter dispersion) can indeed significantly shrink posterior
distributions toward zero. This effect suggests that although prior
variances based on unit information might have desirable behavior with
respect to model determination, they may unintentionally distort the
parameter posterior distributions. We demonstrate that this can affect
the predictive ability of routinely used model averaging approaches in
which information is borrowed across a set of models.

\begin{figure*}
\begin{tabular}{@{}cc@{}}

\includegraphics{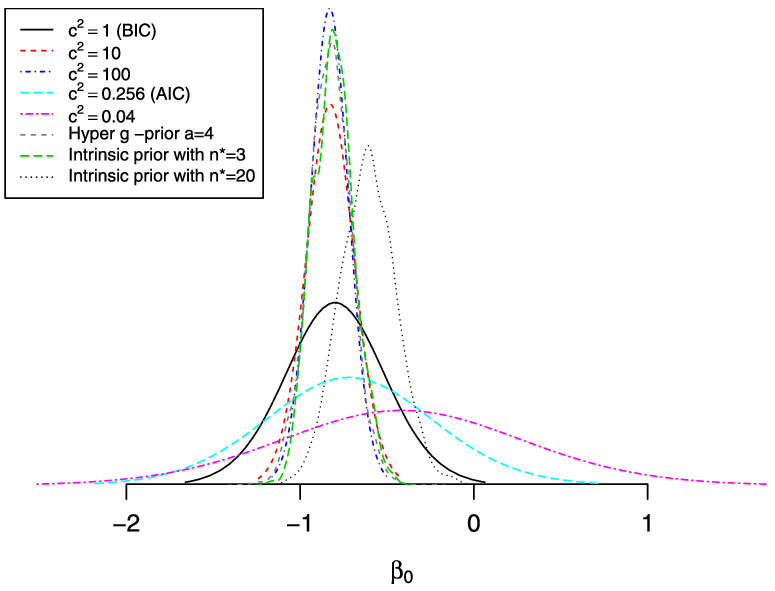}
& \includegraphics{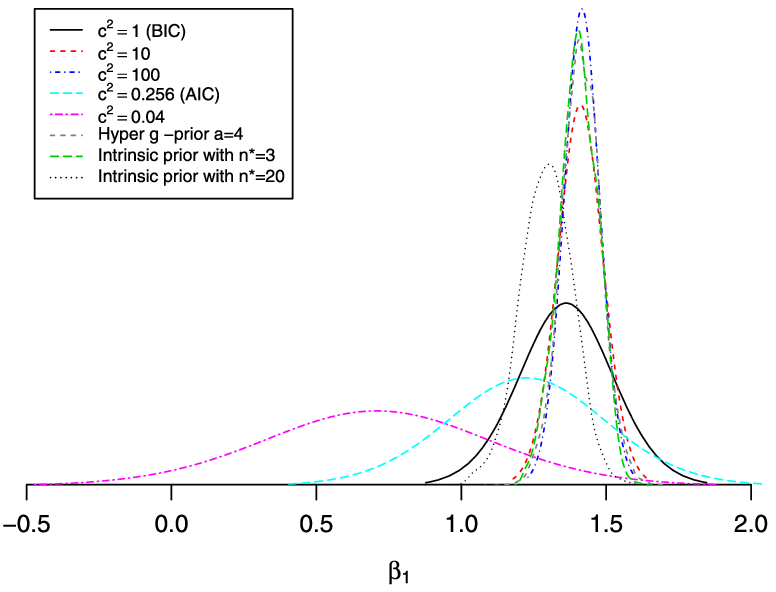}\\
$\beta_0$&$\beta_1$\vspace*{-3pt}
\end{tabular}
\caption{$\!\!\!$Posterior densities of parameters $\beta_0$ and $\beta_1$
under different prior dispersions; $c_m^2\!=\!c^2$ for all models $m$ for
Example~\protect\ref{ex2}.} \label{ex_wind_beta}\vspace*{-4pt}
\end{figure*}

In Section \ref{ex1} we illustrate the effect of Lindley's paradox in a
standard linear regression context emphasizing its dramatic effect on
inference concerning model uncertainty. At the same time, we
demonstrate that if instead of using the standard discrete uniform
prior distribution for $f(m)$ we adopt our proposed adjusted prior
distribution given by (\ref{pmp}) with $p(m)=1$, the prior
distribution for the model parameters can be made highly diffuse in a
way which does not impact strongly on the posterior model
probabilities.

Finally, Section \ref{ex3} investigates the behavior of posterior model
probabilities when substantive prior information about the parameters
is available. We demonstrate through a real data example that the
uniform prior on models may have a significant impact on posterior
model probabilities and we illustrate the advantages of specifying
prior model probabilities that are appropriately adjusted for parameter
prior dispersions.


\subsection{Example 1: A Simple Linear Regression Example}
\label{ex2}

Montgomery, Peck and
Vining (\citeyear{MonPecVin01}) investigated the effect of the logarithm of
wind velocity ($x$), measured in miles per hour, on the production of
electricity from a water mill ($y$), measured in volts, via a linear
regression model of the form
\[
y_i \sim N ( \beta_0 + \beta_1x_i, \sigma^2 ),\quad i=1,\dots,n\vadjust{\goodbreak}
\]
based on $n=25$ data points. We calculate the posterior odds of the
above model, denoted by $m_1$, against the constant model denoted by
$m_0$, adopting the usual conjugate prior specification given by~(\ref{nig}) with zero mean, variance $V_m = c_m^2 n ( \xtx{m} )^{-1}$
and $\alpha=\lambda=10^{-2}$. Since there is a high sample correlation
coefficient of $0.978$ between $y$ and $x$, we expect that $m_1$ will
be a posteriori strongly preferred to $m_0$. Indeed, the posterior
probability of $m_1$ is very close to 1 for values of $ c_m^2$ as large
as $10^{28}$. This behavior provides a source of security with respect
to the choice of $ c_m^2$ and Lindley's paradox, and we use this
example to investigate the effect of $ c_m^2$ on the posterior
densities of $\beta_0$ and $\beta_1$; see Figure \ref{ex_wind_beta}. We
have used values of $ c_m^2$ that represent highly diffuse priors with
$ c_m^2=10$ and $ c_m^2=100$, the unit information prior that
approximates BIC with $ c_m^2=1$, a prior that approximates AIC for
this sample size $c_m^2=(e^2-1)/n=0.256$ and a prior suggested by the
risk inflation criterion (RIC) of Foster and George (\citeyear{FosGeo94}) with $
c_m^2=0.04$; see also George and Foster (\citeyear{GeoFos00}). It is striking that the
resulting posterior densities differ highly in both location and scale.
The danger of misinformation when unit information priors are used was
discussed in detail by Paciorek (\citeyear{Pac06}).

\begin{figure*}
\begin{tabular}{@{}cc@{}}

\includegraphics{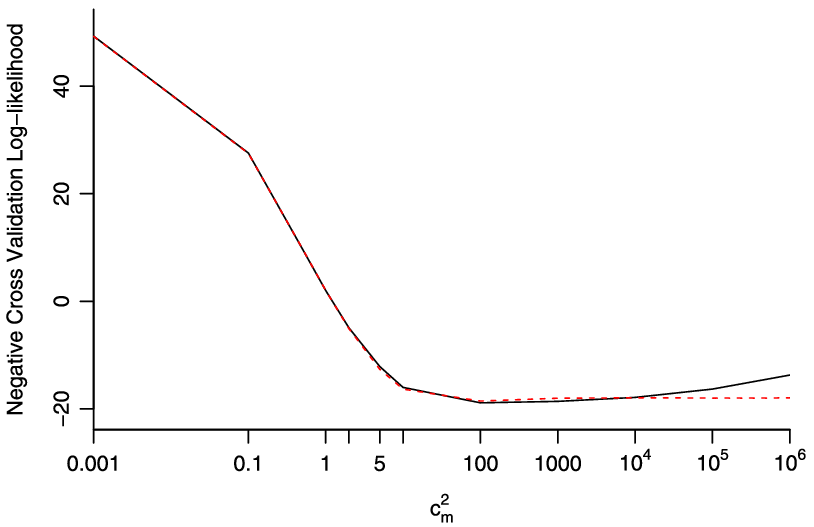}
&\includegraphics{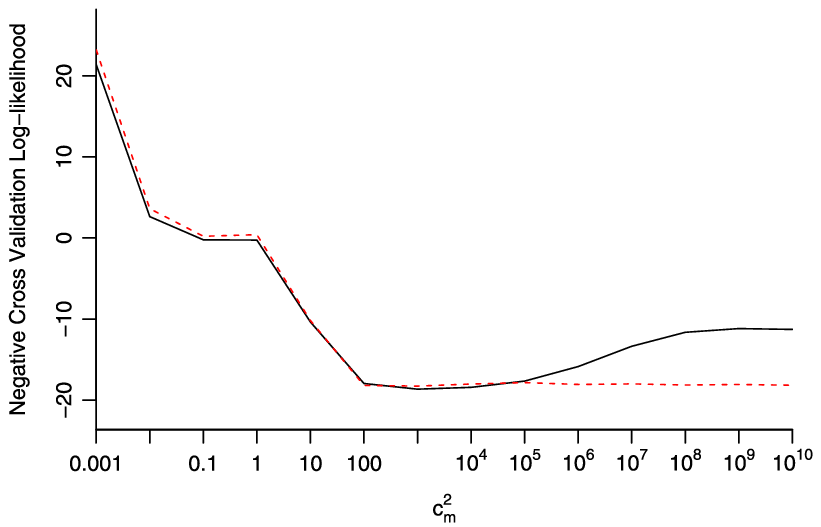}\\
(a) g-prior $( V_m = c_m^2 n ( \mathbf{X}^T_m \mathbf{X}_m
)^{-1} )$. & (b) Independence prior ($V_m = c_m^2
\mathbf{I}_{d_m}$).
\end{tabular}
\caption{Negative cross-validation log-likelihood for two prior
dispersion structures with uniform prior (solid line) and adjusted
prior (dashed line) for Example \protect\ref{ex2}.} \label{ex_wind_ncv}
\end{figure*}

We also investigated how the Zellner and Siow (\citeyear{autokey53}) prior and the
Liang et al. (\citeyear{Liaetal08}) hyper $g$-prior behave in this example. With the
recommended hyperparameter values $2<a\le4$, these priors produced
posterior densities close to the low information $g$-prior with
$c^2_m=100$; see Figure \ref{ex_wind_beta}. The results are quite
robust across this range for $a$ and,~for example, quite large values
of $a$, around 20, are requir\-ed before the level of shrinkage becomes
compara\-ble to the unit information $g$-prior. Hence inferences arising
from the hyper-$g$ prior are quite robust across the recommended range
of hyperparameter values.

Finally, we examined the effect of intrinsic priors on posterior
distributions for model parameters. We adopted the approach of Perez
and Berger (\citeyear{PerBer02}) to construct an intrinsic (or expected posterior)
prior by setting as a baseline prior the $g$-prior with $c^2=100$ and
the null model as a reference. For this simple linear regression model
the minimal training sample has size $n^*=3$. The resulting posterior
distributions of $\beta_0$ and $\beta_1$, also shown in Figure
\ref{ex_wind_beta}, are in close agreement with the baseline
$g$-prior. However, in variable selection problems the minimal
training sample is usually set so that the full model can be estimated.
Hence, the value of $n^*$ could be much higher if more covariates were
available and this would affect the prior variance of the parameters.
As an example, we have calculated the posterior densities of $\beta_0$
and $\beta_1$ when $n^*=20$, also displayed in Figure
\ref{ex_wind_beta}. The effect of the prior densities to the posterior
distributions is dramatic. This nicely illustrates the effect of the
training sample size in intrinsic priors; see the relevant discussion
in Berger and Pericchi (\citeyear{BerPer04}).

We now investigate the effect of prior specification when prediction is
of primary interest. A common way of evaluating predictive performance
is to compute the negative cross-validation score (see Geisser and
Eddy, \citeyear{GeiEdd79}) given by
\[
S = - \sum_{ j = 1 }^n \log f^p(j),\vadjust{\goodbreak}
\]
where
\[
f^p(j) = \sum_{ m \in M} f(m) f ( y_j | \by_{ \setminus j },m )
\]
is the model-averaged predictive density of observation $y_j$ given the
rest of the data $\by_{ \setminus j }$. Lower values of $S$ indicate
greater predictive accuracy. Following Gelfand (\citeyear{Gel96}) we estimate
$f^p(j)$ from an MCMC sample by the inverse of the posterior (over
$m,\bb_m$) mean of the inverse predictive density of observation~$j$.

We generated three additional covariates that have correlation
coefficients $0.99$, $0.97$ and $0.89$ with $x$ and performed the same
model determination exercise. Posterior model probabilities for all
models were calculated for all models under consideration. We used a
$g$-prior with $V_m = c_m^2 n ( \mathbf{X}^T_m \mathbf{X}_m )^{-1}$ and
an independent prior with $V_m = c_m^2 \mathbf{I}_{d_m}$. For the
uniform prior on models combined with the unit information prior
obtained by $ c_m^2=1$, $S$ is far away from the minimum value achieved
for higher values of $ c_m^2$; see Figure \ref{ex_wind_ncv}(a). For $
c_m^2>10^5$, $S$ increases due to the effect of Lindley's paradox
focusing posterior probability on models that are unrealistically
simple. On the other hand, our proposed adjusted prior specification
achieves the maximum predictive ability for any large value of $
c_m^2$; see Figure \ref{ex_wind_ncv}(b). The same exercise was also
repeated for the hyper-$g$ prior for various values of the
hyperparameter $a$. The corresponding negative cross-validation score
was close to the stabilized value of the $g$-prior and it was proven to
be very robust for a wide range of values of $a$. Only for $a$ very
close to $2$, did predictive ability start to deteriorate in a similar
fashion to the $g$-prior.

This simulated data exercise does indicate that predictive ability can
be optimized if highly dispersed prior parameter densities are chosen
together with the adjusted prior over model space. Alternatively, in
this example, the hyper-$g$ family is sufficiently robust to
simultaneously provide a diffuse prior for model parameters, together
with reasonable behavior under model uncertainty.

%

\subsection{Example 2: Simulated Regressions}
\label{ex1}

We now consider the first simulated dataset of Dellaportas, Forster and
Ntzoufras
(\citeyear{DelForNtz02}) based on $n=50$ observations of $15$ standardized independent
normal covariates $X_j,j=1,\ldots,15$, and a response variable $Y$
generated as
%
\begin{equation}Y \sim N (X_4+X_5,~2.5^2).
\label{sampling_scheme_ex1}
\end{equation}
Assuming a conjugate normal inverse gamma prior distribution given by
(\ref{nig}) with zero mean, $V_m = c_m^2 \bsm$ and
$a=\lambda=10^{-2}$, we calculated posterior model probabilities for all
models under consideration. Similar behavior is exhibited either when
$\bsm$ is specified as $\bsm= n ( \xtx{m} )^{-1}$ (described below) or
as $\bsm=\mathbf{I}_{d_m}$.

%

%
%
%

\begin{figure}[t!]
\begin{tabular}{l}
\includegraphics{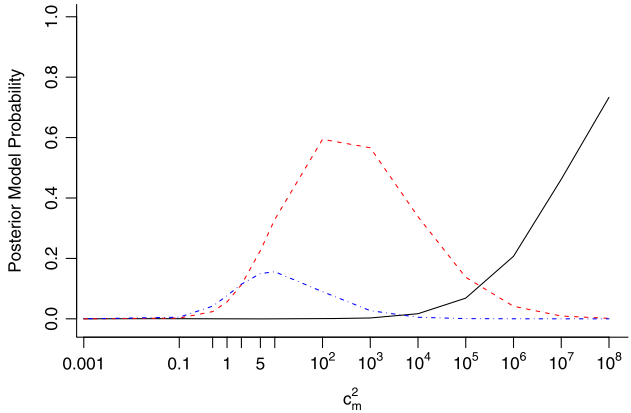}\\
{\fontsize{9.5}{12}{\selectfont{(a) Zellner's $g$-prior with uniform prior on model space.}}}\\[3pt]

\includegraphics{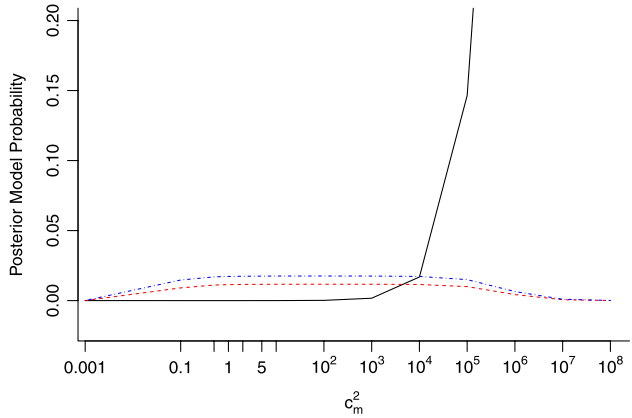}\\
{\fontsize{9.5}{12}{\selectfont{(b) Hyper-$g$ prior with uniform prior on model space.}}}\\[3pt]

\includegraphics{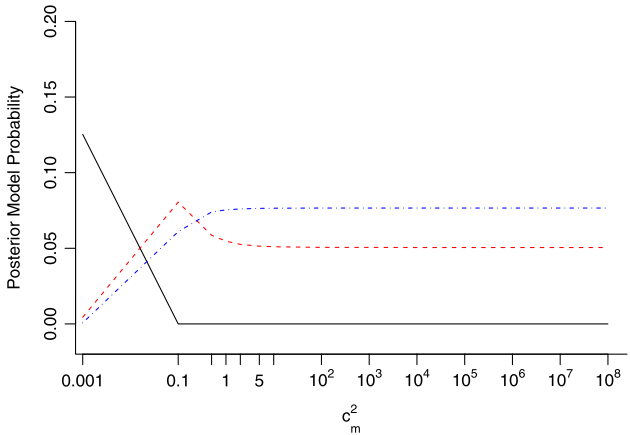}\\
{\fontsize{9.5}{12}{\selectfont{(c) Zellner's $g$-prior with adjusted prior on model
space.}}}\vspace*{-3pt}
\end{tabular}

\caption{Posterior model probabilities under different prior dispersions
for the Dellaportas, Forster and Ntzoufras (\protect\citeyear{DelForNtz02})
dataset of Section \protect\ref{ex1} generated using
(\protect\ref{sampling_scheme_ex1}). Solid line: constant model;
short dashed line: $1+X_4+X_5$ model; long dashed line: $1+X_4+X_5+X_{12}$ model.}
\label{ex1_fig1}
\vspace*{-4pt}
\end{figure}

Figure~\ref{ex1_fig1}(a) and (b), illustrates the behavior of the
posterior model probabilities, under a uniform prior on model space,
of three indicative models. For the parameters we used the $g$-prior
and the hyper-$g$ prior with $c_m^2 = 2n^{-1} /(a-2)$ obtained by
equating the shrinkage proportion $g/(g-1)$ of the $g$-prior with its
prior mean under the hyper-$g$ prior. The effect of Lindley's paradox
is more evident for the $g$-prior where all posterior probabilities are
quite sensitive to the values of $c_m^2$ while the hyper-$g$ prior
demonstrates a remarkable robustness for a wide range of prior
parameter values and only for quite large values of $c_m^2$ which
correspond to values of $a$ close to $2$\vadjust{\goodbreak} is Lindley's paradox
exhibited. We note that the hyper-$g$ prior seems to result in
increased uncertainty on model space resulting in lower posterior model
probabilities for the higher posterior probability models.

\begin{figure}[t!]
\begin{tabular}{l}

\includegraphics{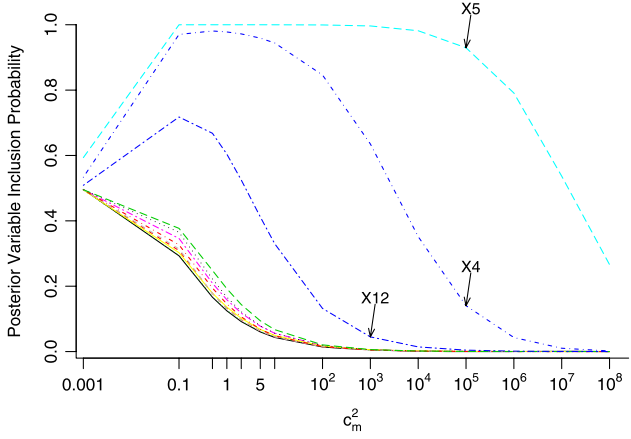}\\
{\fontsize{9.5}{12}{\selectfont{(a) Zellner's $g$-prior with uniform prior on model space.}}}\\[3pt]

\includegraphics{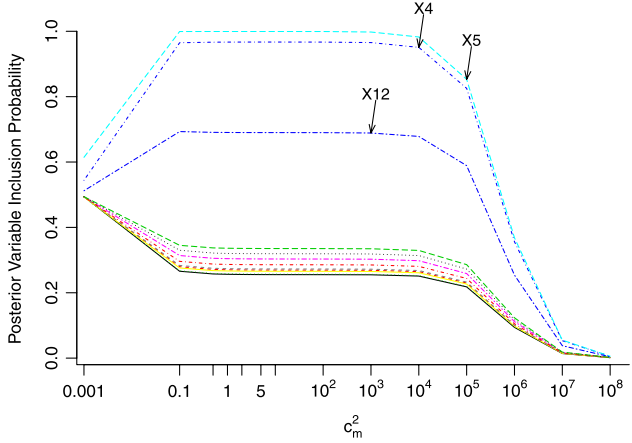}\\
{\fontsize{9.5}{12}{\selectfont{(b) Hyper-$g$ prior with uniform prior on model space.}}} \\[3pt]

\includegraphics{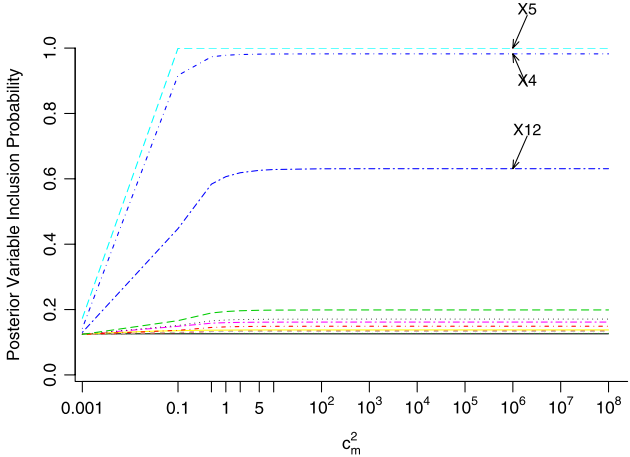}\\
{\fontsize{9.5}{12}{\selectfont{(c) Zellner's $g$-prior with adjusted prior on model space.}}}\vspace*{-3pt}
\end{tabular}
%
\caption{Posterior variable inclusion probabilities under different
prior dispersions for the Dellaportas, Forster and
Ntzoufras (\protect\citeyear{DelForNtz02}) dataset
of Section \protect\ref{ex1} generated using (\protect\ref{sampling_scheme_ex1}).}
\label{ex1_fig2}
\vspace*{-3pt}
\end{figure}

By contrast, using the adjusted prior in Figure~\ref{ex1_fig1}(c)
identifies $1+X_4+X_5+X_{12}$ as the highest probability model for any
value of $ c_m^2>1$. Note that, when $\bsm= n ( \xtx{m} )^{-1}$, $
c_m^2=1$ represents the dispersion induced by the unit information
prior. Similarly, Figure~\ref{ex1_fig2} summarizes the posterior
inclusion probability of each variable $X_j$. Again, for the uniform
prior these probabilities are sensitive to changes in~$c_m^2$ across
its range, whereas the adjusted prior produces stable results for $
c_m^2>1$.

In a more detailed simulation study, we repeated the above analysis by
generating $100$ datasets of the same model. The distribution of the
posterior model probabilities over the $100$ simulated datasets
reinforces the findings of the one-sample based simulation. We also
repeated the above simulation experiment with a more challenging
simulated dataset based on a simulation structure suggested by Nott and
Kohn (\citeyear{NotKoh05}). Each dataset consisted of $n=50$ observations and $p=15$
covariates and one response generated using the following sampling
scheme:
%
\begin{equation}\label{sampling_scheme_new_ex2}
\left\{
\begin{array}{rcl}
X_{j} &\sim& N(0,1) \quad\mbox{for } j=1,\ldots, 10\\
X_{j} &\sim& N( 0.3X_{1}+0.5X_{2}+0.7X_{3}
\\
&&\hphantom{N(}{}+0.9X_{4}+1.1X_{5}, 1
)\\
&&\hspace*{-12pt} \mbox{for } j=11,\ldots, 15\\
Y &\sim& N ( 4 + 2X_{1} - X_{5} + 1.5 X_{7}
\\
&&\hphantom{N(}{}+ X_{11} + 0.5 X_{13},
2.5^2 )
\end{array}
\right\}.
\end{equation}
The general conclusions of this study are in close agreement with
the results obtained above. Further details are available in the
electronic supplement which is available at
\texttt{\href{http://stat-athens.aueb.gr/\textasciitilde jbn/papers/paper24.htm}{http://stat-athens.aueb.gr/}
\href{http://stat-athens.aueb.gr/\textasciitilde jbn/papers/paper24.htm}{\textasciitilde jbn/papers/paper24.htm}}.


\subsection{\texorpdfstring{Example 3: A $3\times2\times4$ Contingency Table Example
with Available Prior Information}{Example 3: A 3 x 2 x 4 Contingency Table Example
with Available Prior Information}}
\label{ex3}

We consider data presented by Knuiman and Speed (\citeyear{KnuSpe88}) to illustrate
how our proposed methodology performs in an example where prior
information for the model parameters is available. The data consist of
$491$ individuals classified in $n$ cells by categorical variables
obesity (O: low, average, high), hypertension (H: yes, no) and alcohol
consumption (A: 1, 1--2, 3--5, 6$+$ drinks per day). We adopt the notation
of the full hierarchical log-linear model used by Dellaportas and
Forster (\citeyear{DelFor99}):
\[
y_i \sim \operatorname{Poisson}
(\lambda_i)\quad\mbox{for }i=1,2,\dots,n,\quad\log(\bolds{\lambda}) =
\mathbf{X}
\bb,
\]
where $\bolds{\lambda} = (\lambda_1, \dots, \lambda_n)^T$,
$\mathbf{X}$
is the $n \times n$ design matrix of the full model, $\bb= ( \bb_j; j
\in{\mathcal V} )$ is an $n \times1$ parameter vector, $\bb_j$\vadjust{\goodbreak} are the
model parameters that correspond to $j$ term and ${\mathcal V}$ is the set
of all terms under consideration. All parameters here are defined using
the sum-to-zero constraints. Dellaportas and Forster (\citeyear{DelFor99}) proposed as
a default prior for parameters of log-linear models
%
\begin{equation}\label{dfprior}
\bolds{\beta}_j
\sim N\pmatrix{ \bolds{\mu}_j,& k_j^2 ( \mathbf{X}_j^T \mathbf{X}_j)^{-1}
 }
\end{equation}
with
$\bolds{\mu}_j$ being a vector of zeros and $k_j^2 = 2 n$ for all $j
\in{\mathcal V} = \{ \varnothing, O, H, A, OH, OA, HA, OHA\}$; we denote this
prior by DF.

In their analysis, Knuiman and Speed (\citeyear{KnuSpe88}) took into account some
prior information available about the parameters $\bb_j$. In
particular, prior to this study information was available indicating
that $\bb_{OHA}$\break and~$\bb_{OA}$ are negligible and only ${\mathcal V} = \{
\varnothing, O, H, A,\allowbreak OH, HA\}$ should be considered. Moreover,
the\break
term~$\bb_{HA}$ is nonzero with a priori estimated effects $
\overline{\bb}_{HA}^T = ( 0.204, -0.088, -0.271 )$; note that the signs
of the prior mean are opposite when compared with reported values of
Knuiman and Speed since we have used a different ordering of the
variable levels.

Knuiman and Speed adopted the prior (\ref{dfprior}) with
$\bolds{\mu}_{HA} = \overline{\bb}_{HA}$ and $\bolds{\mu}_j =
{\bf0}$
for $j \in{\mathcal V} \setminus\{ HA \}$ and prior variance coefficients
$k_{HA}^2=0.05$ and $k_j^2=\infty$ for $j \in\{ \varnothing, O, H, A, OH
\}$. In our data analysis we used $k_j^2 =10^4$ instead of
$k_j^2=\infty$. We denote this prior as KS. We also used a combination
of the DF and KS priors, denoted by KS/DF, modifying slightly the KS
prior so that $k_j^2=2n$ for terms $j \in\{ \varnothing, O, H, A, OH \}$.
Finally, an additional diffuse independence prior, denoted by IND, with
zero prior mean and variance $10^3$ for all model parameters was also
used.

In log-linear models $i(\bb_m)$ depends on $\bb_m$ so to~spe\-cify the
adjusted prior we utilize the prior mean~$\bolds{\mu}_m$ of $\bb_m$
resulting in
\begin{eqnarray*}
&&f(m) \propto p(m) |V_m|^{1/2} | \mathbf{X}_m^T  \operatorname{Diag} (
\bolds{\lambda}_0 ) \mathbf{X}_m |^{1/2} n^{-d_m/2},
\\
&&\quad \bolds{\lambda}_0
= \exp( \mathbf{X}_m \bolds{\mu}_m ),
\end{eqnarray*}
while the prior parameters $p(m)$ were set equal to $\log p(m) =
-\frac{d_m}{2} \log( 2 )$ in line with the DF prior.

\begin{table*}
\tabcolsep=0pt
\caption{Prior and posterior model
probabilities under different parameter and model prior densities for
Example \protect\ref{ex3}} \label{ex2_tab}
\begin{tabular*}{\textwidth}{@{\extracolsep{\fill}}llcccccccccc@{}}
\hline
& \multirow{2}{41pt}{\textbf{Parameter} \textbf{prior}} &\multirow{2}{45pt}{\centering\textbf{Model space} \textbf{prior}} & \multicolumn{4}{c}{\textbf{Prior model probabilities}} & \multicolumn{4}{c@{}}{\textbf{Posterior model probabilities}}\\
\ccline{4-7,9-12}
&  & &  \textbf{O}${}\bolds{+}{}$\textbf{H}${}\bolds{+}{}$\textbf{A} &  \textbf{OH}${}\bolds{+}{}$\textbf{A} & \textbf{O}${}\bolds{+}{}$\textbf{HA} &  \textbf{OH}${}\bolds{+}{}$\textbf{HA} & & \textbf{O}${}\bolds{+}{}$\textbf{H}${}\bolds{+}{}$\textbf{A} &  \textbf{OH}${}\bolds{+}{}$\textbf{A} & \textbf{O}${}\bolds{+}{}$\textbf{HA} &  \textbf{OH}${}\bolds{+}{}$\textbf{HA}\\
\hline
1.&DF & uniform     & 0.25\phantom{0}   & 0.25\phantom{0}   & 0.25\phantom{$0\times ^{1-5}$}              & 0.25\phantom{$0\times ^{1-5}$}                && 0.657    & 0.336     & 0.004 & 0.002
\\
2.&KS & uniform     & 0.25\phantom{0}   & 0.25\phantom{0}   & 0.25\phantom{$0\times ^{1-5}$}              & 0.25\phantom{$0\times ^{1-5}$}                  && 0.075    & 0.000     & 0.923 & 0.002
\\
3.&KS/DF & uniform  & 0.25\phantom{0}   & 0.25\phantom{0}   & 0.25\phantom{$0\times ^{1-5}$}              & 0.25\phantom{$0\times ^{1-5}$}                  && 0.059    & 0.023     & 0.638 &
0.280 \\[5pt]
4.&DF & adjusted    &0.247              & 0.247             & 0.251\phantom{$\times ^{1-5}$}  & 0.255744\phantom{0$^.$}  && 0.677                & 0.317     &
0.004 & 0.002 \\
5.&KS & adjusted    &0.046              & 0.954             & $2.0\times10^{-6}$& $3.3\times 10^{-5}$   && 0.665    & 0.335     & 0.000 & 0.000 \\
6.&KS/DF & adjusted &0.500              & 0.500             & $1.7\times10^{-5}$& $1.7\times 10^{-5}$   && 0.690    & 0.310     & 0.000 & 0.000 \\
7.&IND & adjusted   &0.003              & 0.996             & $3.0\times10^{-6}$& 0.001\phantom{$\times ^{1-5}$}              && 0.690    &0.303      & 0.004 & 0.003 \\
\hline
\end{tabular*}
\end{table*}

Posterior model probabilities (estimated using reversible jump MCMC)
for all prior specifications are presented in Table \ref{ex2_tab}. The
top right panel of the table illustrates the striking effect of
informative parameter priors on posterior model probabilities. The
difficulty of making joint inferences on parameter and model space is
evident by inspecting the sensitivity of model probabilities to
different priors. However, the specification for adjusting the prior
model probabilities has the effect that posterior model probabilities
are robust under all prior specifications.

\section{Conclusion}

There are clearly alternative specifications for the prior model
probabilities $p(m)$ which satisfy (\ref{pmp}), and we do not seek to
justify one over the other. Indeed, choosing model probabilities to
satisfy (\ref{pmp}) may not be appropriate in some situations. Hence,
we do not propose (\ref{pmp}) as a necessary condition for~$f(m)$
although we do believe that there are compelling reasons for
considering such a specification, perhaps as a default or reference
position in the type of situations we have considered in this paper.
What we do argue is that there is nothing sacred about a uniform prior
distribution over models, and hence by implication, about the Bayes
factor. It is completely reasonable to consider specifying $f(m)$ in a
way which takes account of the prior distributions for the model
parameters for individual models. Then, certainly within the contexts
discussed in this paper, as demonstrated by the examples we have
presented, the issues surrounding the role of the prior distribution
for model parameters, in examples with model uncertainty, become much
less significant.

%

\end{document}